\documentclass[aps,prl,twocolumn,showpacs,superscriptaddress]{revtex4}
\usepackage{graphicx}
\begin{document}

%\preprint{}
\title{Resonance in Optimally Electron-Doped Superconductor  
Nd$_{1.85}$Ce$_{0.15}$CuO$_{4-\delta}$}

\author{Jun Zhao}
\affiliation{
Department of Physics and Astronomy, The University of Tennessee, Knoxville, Tennessee 37996-1200, USA
}
\author{Pengcheng Dai}
\email{daip@ornl.gov}
\affiliation{
Department of Physics and Astronomy, The University of Tennessee, Knoxville, Tennessee 37996-1200, USA
}
\affiliation{
Neutron Scattering Sciences Division, Oak Ridge National Laboratory, Oak Ridge, Tennessee 37831-6393, USA}
\author{Shiliang Li}
\affiliation{
Department of Physics and Astronomy, The University of Tennessee, Knoxville, Tennessee 37996-1200, USA
}
\author{Paul G. Freeman}
\affiliation{
Institut Laue-Langevin, 6, rue Jules Horowitz, BP156-38042 Grenoble Cedex 9, France
}
\author{Y. Onose}
\affiliation{Spin Superstructure Project, ERATO, Japan Science and Technology, Tsukuba 305-8562, Japan
}
\author{Y. Tokura}
\affiliation{Spin Superstructure Project, ERATO, Japan Science and Technology, Tsukuba 305-8562, Japan
}
\affiliation{
Department of Applied Physics, University of Tokyo, Tokyo 13-8656, Japan
}

\begin{abstract}
We use inelastic neutron scattering to probe magnetic excitations of an optimally  
electron-doped superconductor Nd$_{1.85}$Ce$_{0.15}$CuO$_{4-\delta}$ 
above and below its superconducting transition temperature $T_c=25$ K.
In addition to gradually opening a spin pseudo gap at the antiferromagnetic ordering wavevector 
${\bf Q}=(1/2,1/2,0)$, the effect of superconductivity is to form a 
resonance centered also at ${\bf Q}=(1/2,1/2,0)$ but at energies above
the spin pseudo gap. The intensity of the resonance develops like a superconducting 
order parameter, similar to those for hole-doped 
superconductors and electron-doped Pr$_{0.88}$LaCe$_{0.12}$CuO$_4$. The resonance is therefore 
a general phenomenon of cuprate superconductors, and must be fundamental
to the mechanism of high-$T_c$ superconductivity.
\end{abstract}

\pacs{74.72.Jt, 61.12.Ld, 75.25.+z}

\maketitle

In conventional Bardeen-Cooper-Schrieffer (BCS) superconductors,  
the superconducting phase forms when electrons are bound into pairs 
with long-range phase coherence through interactions mediated 
by lattice vibrations (phonons) \cite{bcs}. Since high-transition-temperature 
(high-$T_c$) superconductivity arises in copper oxides 
when sufficient holes or electrons are doped into the CuO$_2$ planes of 
their insulating antiferromagnetic (AF) 
parent compounds \cite{kastner}, it is important to determine if 
spin fluctuations play a fundamental role in 
the mechanism of high-$T_c$ superconductivity \cite{scalapino}.  
For hole-doped superconductors, it is now well documented that the spin fluctuations spectrum
forms an `hourglass' dispersion with the most prominent feature,
a collective excitation known as the resonance mode, 
centered at the AF ordering wavevector ${\bf Q}=(1/2,1/2)$  \cite{rossat,fong,dai,woo,stock,fong99,yamada95,mason,tranquada,christensen,vignolle}. 
Although the energy of the mode tracks 
$T_c$ and its intensity behaves like an order parameter below $T_c$ 
for materials such as YBa$_2$Cu$_3$O$_{6+x}$ (YBCO) \cite{rossat,fong,dai,woo,stock}, 
the intensity of the saddle point where the low energy incommensurate spin fluctuations
merge into the commensurate ${\bf Q}=(1/2,1/2)$ point in La$_{2-x}$(Sr,Ba)$_x$CuO$_4$ (LSCO)
displays negligible changes across $T_c$ \cite{tranquada,christensen,vignolle}. Instead, 
the effect of superconductivity in optimally hole-doped
LSCO is to open a spin gap \cite{yamada95} 
and pile density of states along incommensurate wavevectors at energies above the 
spin gap \cite{mason,christensen,vignolle}, and thus appears to be different from
YBCO. 

If the resonance is fundamental to the mechanism of superconductivity, it should
be ubiquitous to all high-$T_c$ superconductors. 
Although the superconductivity-induced 
enhancement at incommensurate wavevectors in LSCO has been argued to be 
comparable to the commensurate resonance in YBCO \cite{tranquadaprb}, 
the intensity gain of the resonance below $T_c$ may not always be 
compensated by opening of a spin gap and spectral weight loss at lower energies.  
For example, the resonance intensity gain in the electron-doped 
Pr$_{0.88}$LaCe$_{0.12}$CuO$_4$ (PLCCO, $T_c=24$ K) 
below $T_c$ is not compensated by spectral weight loss at lower energies \cite{wilson}. 
On the other hand,
 while neutron scattering measurements found a low-temperature spin gap (about 4 meV) in the electron-doped superconductor Nd$_{1.85}$Ce$_{0.15}$CuO$_4$ (NCCO) \cite{tokura,yamada}, there have been no report of the resonance or spectral weight gain at energies above the spin gap below $T_c$.  Therefore, the relationship between the superconducting spin gap
and the resonance is still an open question.

In this Letter, we report the results of inelastic neutron scattering studies of temperature
dependence of the spin fluctuations in an optimally electron-doped NCCO ($T_c=25$ K). We confirm 
the presence of a low-temperature spin (pseudo) gap \cite{yamada} and show that   
the effect of superconductivity also induces a resonance at energies  
similar to electron-doped PLCCO \cite{wilson}. Our results thus demonstrate that the resonance
is an ubiquitous feature of optimally electron-doped superconductors. 
Its intensity gain below $T_c$ in NCCO is due in part to the opening of a spin pseudo gap and 
spectral weight loss at low energies.  This is remarkably similar to the optimally 
hole-doped LSCO \cite{christensen,vignolle}, and 
thus suggesting that the enhancement at incommensurate wavevectors 
below $T_c$ in LSCO has the same microscopic origin
as the commensurate resonance in other high-$T_c$ superconductors. 
  
\begin{figure}[t]
\includegraphics[scale=.40]{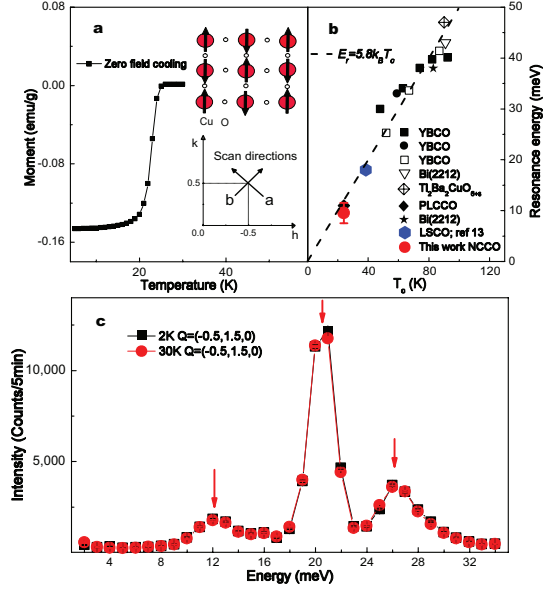}
\caption{a) Schematic diagrams of real and reciprocal space of the CuO$_2$ with the 
transverse and longitudinal scans marked as {\bf a} and {\bf b}, respectively.  
 Magnetic susceptibility measurements of $T_c$. 
 b) Summary of the resonance energy as a function of
 $T_c$ for various hole- and electron-doped superconductors from \cite{wilson}
 with NCCO (this work) and LSCO \cite{christensen} added. 
 c) Energy scans at ${\bf Q}=(-0.5,1.5,0)$ at 2 K and 30 K.  The three 
 CEF levels are marked by arrows \cite{boothroyd}.
 }
\end{figure}

\begin{figure}[t]
\includegraphics[scale=.4]{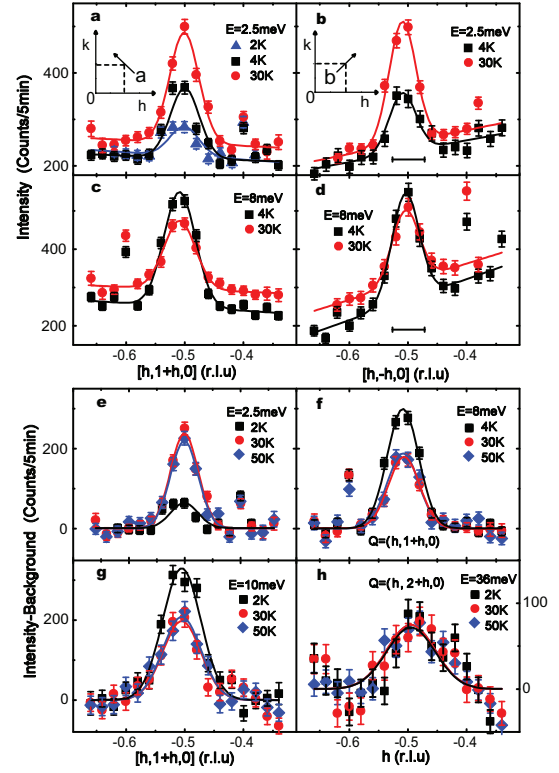}
\caption{Transverse and radial scans through ${\bf Q}=(-0.5,0.5,0)$ for 
a,b) $\hbar\omega=2.5$ meV, and c,d) 8 meV at various temperatures.  Radial scans
in b,d) are instrumental resolution limited (horizontal bars) that gives a minimum 
dynamic spin correlation length $\xi\approx 46$ \AA\ at 2.5 meV.
Transverse scans around ${\bf Q}=(-0.5,0.5,0)$
 with linear background subtracted for 
e) $\hbar\omega=2.5$ meV, f) 8 meV, and g) 10 meV 
at temperature above and below $T_c$. h) The transverse scan
around ${\bf Q}=(-0.5,1.5,0)$ at $\hbar\omega=36$ meV has 
negligible temperature dependence across $T_c$. 
}
\end{figure}

We grew a high quality (mosaicity $<1^\circ$, 3.5 grams) NCCO single crystal using a 
mirror image furnace \cite{kang03}.
Figure 1a plots the magnetic susceptibility measurements showing an onset $T_c$ of 25 K with a 
transition width of 3 K. Our neutron scattering experiments were performed on the 
IN-8 thermal triple-axis spectrometer at the Institute Laue Langevin, Grenoble, France.
We define the wave vector ${\bf Q}$ at $(q_x,q_y,q_z)$ as 
$(h,k,l)=(q_xa/2\pi,q_ya/2\pi,q_zc/2\pi)$ reciprocal lattice units (r.l.u) in
the tetragonal unit cell of NCCO (space group $I4/mmm$, $a=3.95$, and $c=12.07$ \AA ).
For the experiment, the NCCO sample is mounted in the $[h,k,0]$ zone inside a cryostat.  
We chose a focusing Si(111) as 
monochromator and PG(002) as analyzer without collimation. 
The final neutron energy was fixed at $E_f=14.7$ meV with a pyrolytic graphite (PG)
filter in front of the analyzer.  This setup resulted an  
energy resolution of about 1 meV in full-width-half-maximum (FWHM) 
at ${\bf Q}=(-0.5,0.5,0)$. 

\begin{figure}[t]
\includegraphics[scale=.5]{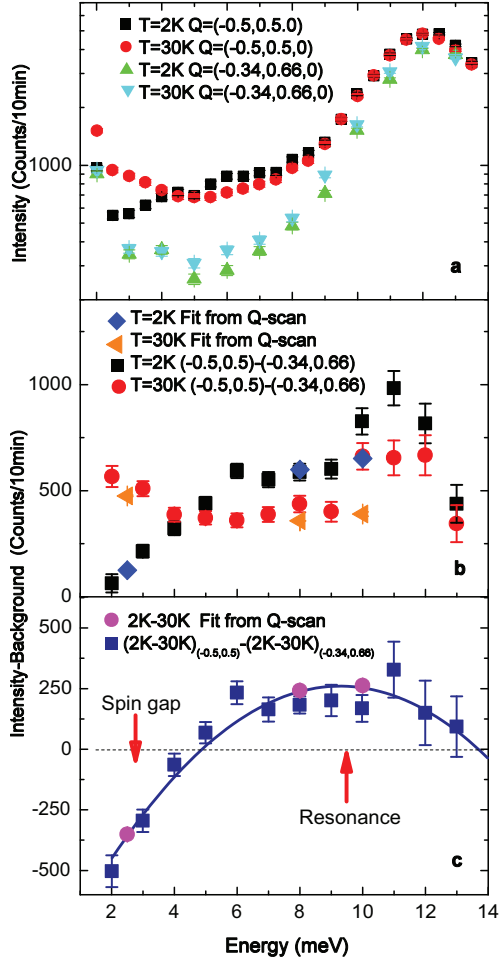}
\caption{a) The temperature dependence of the
scattering at the peak [${\bf Q}=(-0.5,0.5,0)$] 
and background [${\bf Q}=(-0.34,0.66,0)$] 
positions below and above $T_c$.  Note the intensity is 
plotted in log-scale to display the large intensity
difference between the Nd$^{3+}$ CEF level at $\hbar\omega=12$ meV
and Cu$^{2+}$ spin fluctuations centered at 
${\bf Q}=(-0.5,0.5,0)$ for energies between 2 and 10 meV.
b) Background subtracted magnetic scattering at 
${\bf Q}=(-0.5,0.5,0)$ below and above $T_c$.  
The data are cross checked by constant-energy scans in Fig. 2.
c) The temperature difference plot showing the resonance
at $E_r=9.5\pm2$ meV.  The large error is due to the uncertainty in
obtaining Cu$^{2+}$ magnetic signal above 10 meV.
}
\end{figure}

To understand the effect of superconductivity on the Cu$^{2+}$ spin fluctuations, we must first 
determine the temperature dependence of the magnetic excitations from Nd$^{3+}$ crystal 
electric field (CEF) levels in NCCO. For Nd ions in the tetragonal 
NCCO crystal structure, the three lowest energy CEF magnetic excitations are at 
$\hbar\omega=12.2\pm0.3$ meV, $20.3\pm0.1$ meV, and $26.5\pm0.3$ meV \cite{boothroyd}. 
Our energy scans at ${\bf Q}=(-0.5,1.5,0)$ 
confirm these results and show that the intensities of these CEF levels have small temperature
dependence between 2 K and 30 K (Figure 1c).

Figure 2 summarizes the transverse and longitudinal 
${\bf Q}$-scans around $(-0.5,0.5,0)$ at different energy transfers and temperatures.     
Consistent with earlier results on NCCO \cite{yamada} and PLCCO \cite{wilson,wilsonprb},
the scattering is commensurate and centered at ${\bf Q}=(-0.5,0.5,0)$ for 
all energies probed. Figures 2a-d show the raw data (with scan directions marked) below and above $T_c$ at $\hbar\omega=2.5$, 8 meV.
At $T=30$ K ($T_c+5$ K), the magnetic scattering above the linear backgrounds decreases slightly with increasing
energy from 2.5 meV to 8 meV (Figs. 2e and 2f).  On cooling to below $T_c$, the peak intensity 
is drastically suppressed for $\hbar\omega=2.5$ meV (Figs. 2a and 2b), and 
it increases for $\hbar\omega=8$ meV (Figs. 2c and 2d).  Figures 2e-g show background subtracted 
transverse scans at various energies. It is immediately clear that cooling 
below $T_c$ suppresses the ${\bf Q}=(-0.5,0.5,0)$ peak at $\hbar\omega=2.5$ meV but enhances scattering
at $\hbar\omega=8$ and 10 meV.  On the other hand, magnetic scattering at $\hbar\omega=36$ meV changes
negligibly from 2 K to 50 K (Fig. 2h).

Figures 3a and 3b show energy scans at the signal [${\bf Q}=(-0.5,0.5,0)$] and 
background [${\bf Q}=(-0.34,0.66,0)$] positions above and below $T_c$. 
Although the large Nd$^{3+}$ CEF level dominated the magnetic
scattering at $\hbar\omega=12$ meV \cite{boothroyd}, 
one can still see clear Cu$^{2+}$ spin fluctuations centered at $(-0.5,0.5,0)$ 
for energies between 2 and 10 meV.  In the normal state,
the magnetic scattering decreases with increasing energy, consistent with  
${\bf Q}$-scans at $\hbar\omega=2.5$, 8, and 10 meV (Figs. 2e-g). 
In the superconducting state, the low-energy 
spin fluctuations at ${\bf Q}=(-0.5,0.5,0)$ are suppressed for $\hbar\omega\le 4$ meV and
there is a clear scattering intensity gain for $6\le\hbar\omega\le 10$ meV.
The contrast between the normal and superconducting states becomes more 
obvious when changes in background scattering are taken into account (Fig. 3b).  
The large Nd$^{3+}$ CEF scattering between 
$10<\hbar\omega<33$ meV (Fig. 1c) overwhelmed Cu$^{2+}$ magnetism.
The background corrected difference 
plot between the superconducting and normal states shows a resonance at
$\hbar\omega=9.5\pm2$ meV, similar to that for PLCCO \cite{wilson}.

\begin{figure}[t]
\includegraphics[scale=.5]{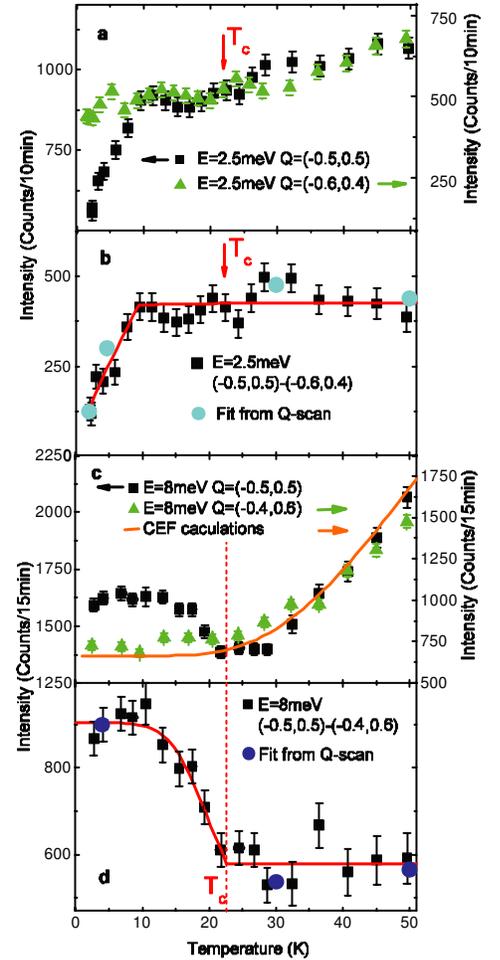}
\caption{Temperature dependence of the scattering at 
$\hbar\omega=2.5$, and 8 meV. a) The raw data at the signal
[${\bf Q}=(-0.5,0.5,0)$] and background [${\bf Q}=(-0.6,0.4,0)$] positions.
b) The background subtracted magnetic scattering at $\hbar\omega=2.5$ meV
shows no anomaly cross $T_c$ but drops dramatically below 9 K.  The data from
the fitted ${\bf Q}$-scans are shown as circles.
c) Temperature dependent data for $\hbar\omega=8$ meV with background at 
${\bf Q}=(-0.4,0.6,0)$, a resonance
coupled to $T_c$ like an order parameter is clearly seen in the
background subtracted data in d).  The estimated temperature
dependence of the Nd$^{3+}$ CEF level at 8 meV 
(from 12 meV to 20 meV) is shown as solid line in c) \cite{boothroyd}. 
}
\end{figure}

To determine if the low temperature spin fluctuations' suppression 
below 4 meV and enhancement between 6 to 10 meV are indeed associated with the opening of a
superconducting gap below $T_c$ as in the tunneling experiments \cite{shan}, we 
carefully measured the temperature dependent scattering at the peak [${\bf Q}=(-0.5,0.5,0)$] and
background positions for $\hbar\omega=2.5$ and 8 meV.  
From previous low-energy inelastic neutron scattering work on NCCO \cite{yamada}, 
we know that the spin gap in NCCO opens gradually with decreasing temperature until it reaches
to about 4 meV at 2 K.  While peak intensity 
in the ${\bf Q}$-scans at $\hbar\omega=2.5$ meV show a 
clear low temperature suppression, there is still a peak 
present at ${\bf Q}=(-0.5,0.5,0)$ even at 2 K.  Therefore, optimally electron-doped 
NCCO does not have a clean spin gap as in the case of the optimally hole-doped
LSCO \cite{yamada95}.  
The temperature dependence of the scattering at the peak and background positions  
(Figs. 4a and 4b) reveals that the intensity suppression at 
$\hbar\omega=2.5$ meV does not happen at $T_c$ but at 9 K ($T_c-16$ K).  While this 
result confirms the earlier report \cite{yamada}, it also suggests that the gradual 
opening of the (pseudo) spin gap is not directly related the temperature dependence of
the superconducting gap which is BCS-like \cite{shan} and becomes essentially fully opened
with $2\Delta\approx 7$ meV below 12 K (50\% of $T_c$).

On the other hand, the temperature dependence of the scattering 
at $\hbar\omega=8$ meV is clearly coupled to the occurrence of superconductivity. 
With increasing temperature, the scattering at ${\bf Q}=(-0.5,0.5,0)$ first decreases
like an order parameter, showing a kink at $T_c$, and then increases again above 30 K.
It turns out that the large intensity rise above 30 K at $\hbar\omega=8$ meV 
is due to the CEF transition from 12 meV to 20 meV as the 12 meV state is being populated with
increasing temperature (Fig. 4c) \cite{boothroyd}. As the CEF levels are 
weakly {\bf Q}-dependent, the large intensity increase above 30 K is also 
seen in the background (Fig. 4c).  The difference between signal and background shows a clear
order-parameter-like temperature dependence of the resonance, remarkably similar to the
temperature dependence of the resonance in PLCCO \cite{wilson} 
and hole-doped superconductors \cite{rossat,fong,dai,woo,stock,fong99}.

The discovery of the resonance in another class of electron-doped superconductors
suggests that the mode is a general phenomenon of electron-doped superconductors independent of
their differences in rare-earth substitutions \cite{tokura}.  
For hole-doped LSCO \cite{yamada95,mason,tranquada,christensen,vignolle}, the intensity enhancement in 
spin susceptibility above the spin-gap energy has been characterized as the magnetic coherence effect \cite{mason,tranquadaprb}. The observation of the susceptibility enhancement at energies 
($6\le\hbar\omega\le13$ meV)
just above the spin pseudo gap energy of 4 meV 
in NCCO is consistent with this picture, although the temperature dependence of 
the spin pseudo gap in NCCO behaves rather differently from those in LSCO \cite{yamada,yamada95}.
In our search for the excitations responsible for electron pairing and 
high-$T_c$ superconductivity, one
of the arguments against the relevance of the resonance has been the inability to observe 
superconductivity-induced commensurate resonance  
in LSCO \cite{yamada95,mason,tranquada,christensen,vignolle}. If the resonance is 
a phenomenon associated with the opening of a superconducting gap and the subsequent
local susceptibility enhancement, it is natural to regard the susceptibility gain in both 
NCCO and LSCO as the resonance.  Adding these two points to the universal $E_r=5.8k_BT_c$ 
plot in Fig. 1b suggests that 
 while the resonance energy itself is intimately related to
$T_c$, other details such as the spin gap, commensurability, and hourglass dispersion 
found in different materials may not be fundamental to the superconductivity.

For hole-doped superconductors, the hourglass dispersion has been interpreted either 
as the signature of ``stripes'' where doped holes are 
phase separated from the Mott-like AF background \cite{kivelson,zaanen,vojta},
 or as a bound state (spin exciton) within
the gap formed in the non-interacting particle-hole continuum of a Fermi-liquid \cite{ermin05,eschrig}.
Although the resonance in PLCCO has been interpreted as an over damped spin
exciton \cite{ermin07}, it remains a challenge to understand how the resonance can arise both from 
NCCO which has a spin pseudo gap and from the gapless PLCCO \cite{frank}.

We thank Stephen Wilson and Jeff Lynn for earlier experiments on NCCO at NIST.
This work is supported by the US DOE BES under contract No. DE-FG02-05ER46202.  ORNL is supported by US DOE 
DE-AC05-00OR22725 with UT/Battelle LLC.

%\bibliography{NoEndingPoint}

\end{document}